\documentclass[twocolumn,floats,aps]
{revtex4}\usepackage{times,amsmath,graphics,psfrag,latexsym}
\begin{document}
\title{Energy Loss from Reconnection with a Vortex Mesh}
\author{I.H. Neumann and R.J. Zieve}
\affiliation{Physics Department, UC Davis}
\begin{abstract} 
Experiments in superfluid ${}^4$He show that at low temperatures, energy
dissipation from moving vortices is many orders of magnitude larger than
expected from mutual friction. Here we investigate other mechanisms for energy
loss by a computational study of a vortex that moves through and
reconnects with a mesh of small vortices pinned to the container wall. We find
that such reconnections enhance energy loss from the moving vortex by a factor
of up to 100 beyond that with no mesh.  The enhancement occurs through two
different mechanisms, both involving the Kelvin oscillations generated along the
vortex by the reconnections.  At relatively high temperatures the Kelvin waves
increase the vortex motion, leading to more energy loss through mutual
friction.  As the temperature decreases, the vortex oscillations generate
additional reconnection events between the moving vortex and the wall, which
decrease the energy of the moving vortex by transfering portions of its length
to the pinned mesh on the wall.

\end{abstract}
\maketitle

\section*{Introduction}

Energy transfer in superfluids is closely tied to the quantized vortex lines
present in any non-trivial flow \cite{Donnelly}.  In a superfluid the
curl of the velocity field vanishes except along isolated vortex cores, and its
values along the cores lead to quantized circulation within the superfluid.
Superflow, with no energy loss, occurs only at sufficiently low speeds.  Higher
velocities lead to creation of vortices, which extracts energy from the flow,
and then to further energy loss as those vortices move.

Early measurements of energy dissipation in superfluid ${}^4$He include 
second-sound propagation through vortex arrays in uniformly rotating resonators
\cite{HV} and electrical detection of individual charged vortex rings moving
through an otherwise stationary fluid \cite{RR}.  These experiments show that
dissipation depends strongly on the fluid's temperature as well as on the
vortex velocity.  At relatively high temperatures, above 1 K, thermal
excitations behave like a normal liquid of density $\rho_n$, coexisting
with a superfluid of density $\rho_s$.  The sum $\rho_n+\rho_s$ gives the total
fluid density.  Since a vortex core is an excitation but the flow about the
core is superfluid, vortices couple the two fluids through a force known as
mutual friction.  Experimentally, the mutual friction force per unit length of
vortex is \cite{BD}
\begin{equation} 
{\bf F}_f=\alpha \kappa\rho_s({\bf v}_n-{\bf v}_L). 
\label{e:fricforce} 
\end{equation}  
Here $\kappa$ is the circulation around the moving vortex, $\alpha$ is an
experimentally determined friction parameter, and ${\bf v}_n$ and ${\bf v}_L$
are the velocities of the normal fluid and the vortex, respectively. In general
the mutual friction force has an additional component perpendicular to the
vortex line's velocity, which we neglect because its magnitude is much smaller
\cite{HV}. 

Well below 1 K, the normal fluid fraction becomes negligible and transfer of
energy from the superfluid to normal component through mutual friction is no
longer significant.  Recent interest in low-temperature dissipation mechanisms
has centered around the problem of superfluid turbulence, where groups of
vortices form tangles which highly resemble classical turbulence.  Many length
scales are represented in the tangles, depending on the curvatures of different
vortex segments, and the velocity field changes drastically on short length
scales near any individual vortex core.  Above 1 K, turbulence generated on a
macroscopic length scale, as by rotating blades \cite{Tabeling}, a towed grid
\cite{Smith}, or spin-down of a rotating container \cite{Walmsley07}, shows the
same energy spectrum and time decay as turbulence in classical fluids. This can
be explained as classical turbulence in the normal component, coupled to the
superfluid through mutual friction.  As temperature decreases, the decay rate
for vortex line length drops abruptly near 0.8 K \cite{Walmsley07}, although the
functional form remains unchanged.  This indicates a change in the mechanisms
for dissipating energy or for transfering energy to smaller length scales.  When
the turbulence is created directly by injecting microscopic vortex rings, the
functional form itself changes at low temperatures \cite{Walmsley08, Tsubota}. A
similar effect is seen with inhomogeneous tangles in superfluid ${}^3$He-B
\cite{Bradley06}. Current ideas for the low-temperature dissipation mechanism
center on vortex reconnections and on phonon emission by Kelvin waves along
vortex lines \cite{Vinen01, Charalambousreview}.  To explore such ideas, further
experiments and simulations are needed.

\begin{figure}[htb]
\begin{center}
\scalebox{.35}{\includegraphics{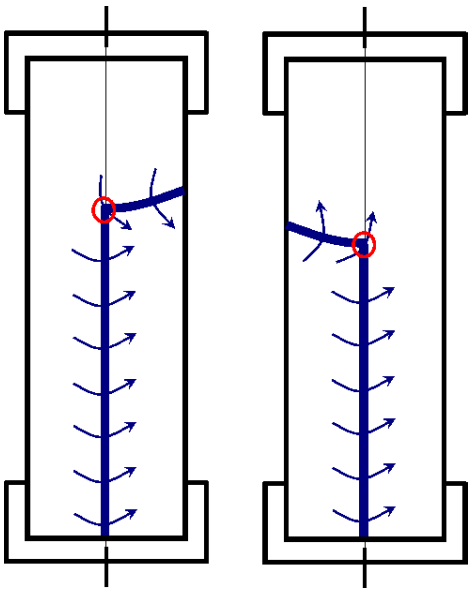}}
\caption{\small A vortex partially pinned to a wire, with core in
blue, at two different times.  The point where the vortex detaches
from the wire is marked with a red circle.  The unpinned portion
precesses about the cylindrical cell.  Arrows represent the velocity field near
the vortex core.}
\label{f:cell}
\end{center}
\end{figure}

Our present work is motivated by measurements in a unique single-vortex
geometry \cite{Donev}.  A fine wire, stretched along the axis of a
cylinder filled with superfluid, traps a vortex.  If the vortex partially
depins from the wire and continues from the wire to the curved cell wall, as
illustrated in Figure \ref{f:cell}, the
free portion of the vortex moves through the cell at the local superfluid
velocity.  Since the velocity field is determined primarily from the location
of the vortex core, the free end precesses about the wire.  During the
precession, the portion of the vortex that remains trapped about the wire
decreases in length, which indicates that the moving vortex dissipates
energy.  However, the mechanism is unclear; at the lowest temperatures, near
320 mK, the dissipation is many orders of magnitude larger than could occur
through mutual friction.  We note that the vortex velocity in this measurement
is very slow, roughly 0.003 cm/s as opposed to 20-120 cm/s in \cite{RR}. 
Other measurements do find some velocity dependence in the mutual friction
coefficient $\alpha$ at high temperature, but this effect is negligible below
1 K \cite{Swanson}. Furthermore, the rate of energy loss through mutual
friction should be proportional to $\alpha$, but in fact the measured energy
loss has far less temperature dependence than $\alpha$.  This suggests that a
different mechanism is at work.

The likely source of the observed dissipation is a vortex-wall interaction.
Since the moving vortex terminates on the container wall, it can interact with
the surface in addition to undergoing vortex-fluid interactions along its
length. The exact mechanism of the vortex-wall interaction is unknown.
Possibilities suggested previously include viscous drag through a normal-fluid
Ekman layer near the surface and a force from vortex line tension
\cite{Adams}.  Since neither of these completely fits the experimental results
\cite{expt}, we here investigate another possibility: an interaction with
microscopic vortex segments pinned to the cell wall.  As the moving vortex line
sweeps along the wall, it encounters pinned vortices.  Closely approaching
vortices undergo reconnections, and the moving vortex may gradually shrink
if it leaves pieces of itself behind on the cell wall. In this paper we
describe numerical simulations of the process.  In common with the recent
efforts on superfluid turbulence, it proves to depend heavily on interplay
between vortex reconnections and the Kelvin waves they generate.  

\section*{Computational Model}

We follow the method originally used by Schwarz for modeling
vortex dynamics \cite{Schwarz}.  As described elsewhere \cite{simSamuels,
simTsubota, simAarts, Aartsthesis, simqucl}, we treat the vortex cores as
massless and thin; that is, the core diameter is small compared to the
curvature along the core. The equation governing the motion of superfluid
vortices is essentially the Euler equation for incompressible fluid flow, with
an additional term for mutual friction between the vortices and normal fluid. 
Physically, in the absence of friction, the vortex cores move at the local
superfluid velocity. In our code, we specify a vortex by a series of points
along its core and update each point based on the fluid velocity at that
location, calculated in the local induction approximation.
Details of our algorithm 
appear in \cite{lukesim}. We use a fourth-order Runge-Kutta-Felberg (RKF) method
with a variable time step for solving the ordinary differential equation.  We
determine the step size by requiring third-order and fourth-order RKF
calculations to agree to 0.1\%.

We calculate the superfluid velocity field from the positions of the vortex
cores, adding an additional zero-curl field to ensure that the boundary
condition at the wall is satisfied: namely, that the velocity component
perpendicular to the wall must vanish at the wall. In the present work we
compute this boundary field approximately using image vortices, as described
below.  

Following \cite{Schwarz}, we include a mutual friction term based on Equation
\ref{e:fricforce}.  The friction term damps out oscillations of the vortex
lines, keeping the simulations stable.  For this purpose we use values of
$\alpha$ from 0.001 to 0.1, which are much larger than the actual friction
coefficient $\alpha$ = $5\times 10^{-10}$ for superfluid helium at our
experimental temperature \cite{coeff}.  The temperatures corresponding to
the friction coefficients in our simulations are 850 mK to 1.6 K. 

In the present simulations, we omit an explicit calculation of the contributions
to the velocity field from distant vortex cores.  Instead we incorporate their
influence through reconnections.  The velocity field due to a vortex falls off
quickly with distance, so it is reasonable to consider only close neighbors. 
Furthermore, there is some averaging of the contributions from more distant
vortices.  Reconnections occur because, for most vortex orientations, two
closely approaching vortices attract each other with increasing strength
\cite{Schwarz, recon}.  They form cusps, which get drawn out ever larger. 
Simulations typically break down in this regime, on account of the rapidly
varying velocity fields that nearby vortices feel from each other. Experiments
show reconnections of the two vortices, with the new formations then moving
apart from each other \cite{Paoletti}. Following \cite{Schwarz}, we reconnect
any two vortices that approach each other closer than a cutoff distance.  To
avoid recurring reconnections between the same pair of vortices, a vortex
segment that has just undergone a reconnection cannot reconnect again for a
brief period.  This allows the newly reconnected segments to separate beyond the
reconnection distance.

We also allow reconnections between a vortex and the cell wall.  Since the
effect of a surface can be treated with image vortices, a vortex which closely
approaches the wall undergoes a cusp formation similar to that in vortex-vortex
interactions.  When the vortex touches the wall, two new vortex endpoints form
and proceed to move apart from each other.

\begin{figure}[htb]
\begin{center}
\scalebox{.4}{\includegraphics{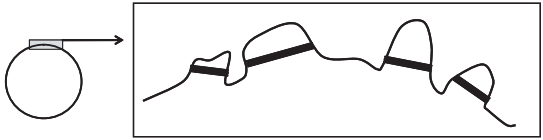}}
\caption{\small An exaggerated schematic of the cylinder wall's roughness. Mesh 
vortices pin to protrusions and are pulled taut between them.}
\label{f:WallMesh}
\end{center}
\end{figure}

We simulate the motion of a vortex that runs from the cylinder axis to its
curved wall.  We assume that the end that reaches the axis continues as a
straight, stationary vortex along the axis, analogous to the experimental 
situation described above with a vortex line partially pinned to a central wire
\cite{Donev}. In addition, we consider a mesh of microscopic vortices attached
to the surface of the cell. Because of the extremely small coherence length in
superfluid ${}^4$He, such pinned vortices appear upon cooling through the
superfluid transition temperature, where the energy barrier to vortex creation
is small \cite{Awschalom}.  They can also appear when the cryostat rotates, as
it does in the experiment.  Rotating creates an array of straight vortex lines
parallel to the rotation axis. Once rotation ceases, these vortices annihilate
at the cell wall, which is rough on the scale of a superfluid vortex.  Fragments
of the vortices could remain pinned between small protrusions along the cell's
walls, as illustrated in Figure \ref{f:WallMesh}.  We examine how such a mesh of
microscopic vortices affects the macroscopic vortex running from the wire to the
cell wall. 

We do not attempt to model the cell's surface roughness explicitly; instead we
define certain points on the cylinder's wall as the endpoints of pinned vortices.
From previous work, we know that the only stable macroscopic vortices pinned at
both ends to the cylinder wall have significant horizontal components and span
much of the cell \cite{lukesim}.  However, the mesh we envision here consists of
much shorter vortices.  On their length scale the wall does
not resemble a smooth cylindrical surface and the earlier calculation does not apply. For
most of our work here, we assume that the mesh vortices are stretched taut
between their two endpoints.  Figure \ref{f:WallMesh} shows our picture
of vortices pinned to the wall.  Alternatively, in a few calculations we use
semicircular rather than straight-line vortices.  In generating the pinned vortices, we can vary their
number, length, and orientation. 

To set up the initial vortex mesh, we select a portion of the cell wall on which
the vortices can lie.  We then choose a random point within this region as
one end of a mesh vortex.  We choose the second endpoint relative to the first
one.  The orientation and length of the vortex each can be fixed or can be
chosen at random within a selected range.  We then repeat the process to
generate a target number of mesh vortices.  One final step is needed for
simulations with random orientations.  Our algorithm requires only the first
endpoint to lie within the selected region, so all vortices that cross the edge
of the mesh region initially point outwards. We eliminate this bias by flipping
the orientation of each vortex with 50\% probability.  

Our RKF algorithm updates only the position of the vortex line which stretches
from the cylinder axis to the wall.  The mesh vortices and the vortex portion
running along the cylinder axis remain stationary.  

As noted above, we use image vortices to satisfy the boundary conditions for
the velocity field approximately. For the central vortex, we extend it along
the cylinder axis to negative infinity, and also continue the other end from
the curved wall to infinity as a straight, radial vortex. For the mesh
vortices, we note that at an infinite plane boundary, image vortices produce an
exact solution to Laplace's equation that satisfies the boundary condition. 
Since the mesh vortices are extremely short compared to the cylinder radius,
the cylinder wall can be treated as nearly flat, apart from the wall
roughness.  To adjust for its curvature, we invert each mesh vortex in the
cylinder, so that with its image it becomes a vortex ring. To the extent that
these methods do not perfectly meet the boundary conditions, we can picture
deformations in the cylinder which do make its surface perpendicular to
the velocity field.  These deformations are a type of wall roughness,
which we already assume exists when we define the vortex mesh.

As the moving vortex sweeps along the wall, it encounters mesh vortices and
undergoes reconnections.  In principle, the moving vortex could also bend enough
that it reconnects with itself, but in practice this does not happen for the
present geometry. When a
reconnection occurs, one endpoint of the mesh vortex becomes the wall terminus
of the moving vortex while the original wall terminus becomes a mesh vortex
endpoint. We then redefine the mesh vortex using the new endpoints and assuming
it runs directly between them.  Straightening the mesh vortex makes sense
because the actual motion after a reconnection involves oscillations stemming
from cusp formation at the reconnection point \cite{Schwarz, recon, Paoletti}.
The vortex gradually loses energy through these oscillations, and a pinned
vortex would eventually become straight.  Although we do not calculate the
oscillations of the pinned vortex after each reconnection, in the corresponding
experiment the pinned vortices do have plenty of time to settle before the
moving vortex interacts with them, so they are likely to be nearly straight.

\begin{figure}[tbh]
\begin{center}
\scalebox{.74}{\includegraphics{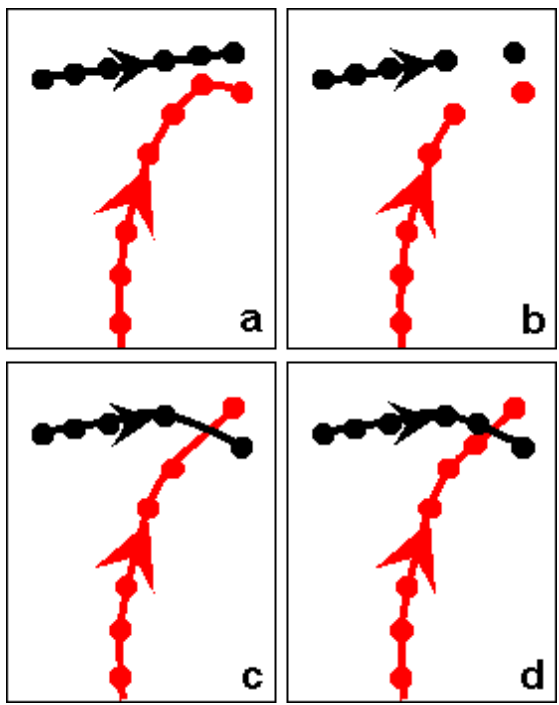}}
\caption{\small Reconnection between two vortices.  Vortices
approach each other (a).  At least one point on each vortex is removed (b).
Remaining points on either side of the removed point are reconnected, swapping
between vortices (c).  If the new straight segments are too long, additional
points are inserted (d).
}
\label{f:reconalg}
\end{center}
\end{figure}

More detailed simulations \cite{Leadbeater} show that vortex
reconnections release energy into sound waves.  However, the energy loss
is so small that thousands of reconnections per second would be required
to match our experimental observations.  The present work investigates less
direct energy consequences of reconnections.  Our reconnection algorithm
identifies the points of closest approach of two vortices, then
moves out along each vortex a distance comparable to the vortex separation.
We delete the segments near the reconnection points and replace them with
segments crossing between the original vortices, as shown in Figure
\ref{f:reconalg}. With the numbers we use for the various lengths, on average
the vortex length actually increases slightly through reconnections.  Thus we
are not trivially introducing energy loss to our simulation through the numerics
of our reconnections.

In our previous calculations on nearly stationary vortices, we subtracted the
velocity tangent to the vortex core before updating the configuration. Here we
keep the tangential component.  Points along the vortex line occasionally bunch up
or separate too far, so we delete or add points as appropriate to keep
neighboring point spacings fairly uniform.  Near the cell wall the point spacing
must be smaller than the length of the mesh vortices.  However, maintaining such
a small spacing along the entire moving vortex would require vast numbers of
data points, with correspondingly long computation times.  Since the moving
vortex has a much larger radius of curvature once it leaves the wall region, a
lower density of points is adequate.  We increase the point spacing by an order
of magnitude far from the cell wall.

\begin{figure}[htb]
\begin{center}
\scalebox{.33}{\includegraphics{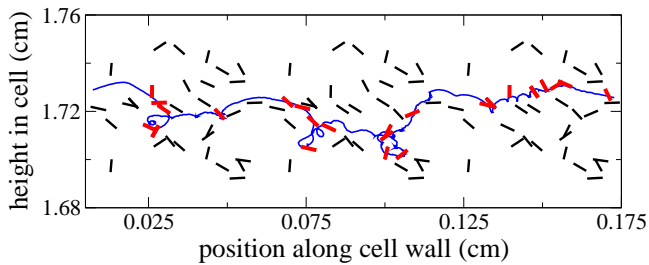}}
\caption{\small During one run with $\alpha=0.01$, the wall
terminus of the moving vortex follows the path along the cylinder wall
indicated by the long blue curving trace.  The mesh vortices it reconnects
with are thick and red, while the other mesh vortices are thinner
and black.}
\label{f:path}
\end{center}
\end{figure}

Any reconnection produces oscillations along the new vortex lines \cite{TT,
Kelvin}. Immediately after a reconnection sharp kinks in the moving vortex
appear close to the wall. As the vortex line tension acts on the kinked
regions,  characteristic Kelvin oscillations ensue.  Mutual friction damps out
the Kelvin oscillations from a single reconnection, but a steady rate of
reconnections can maintain the oscillatory motion.  The oscillations persist
along the entire vortex, with most of the additional motion vertical.  For our
moving vortex, the oscillations increase the distance the vortex travels over
the cell wall, enabling it to encounter additional mesh vortices and undergo
further reconnections.  Figure \ref{f:path} illustrates the path of the vortex
endpoint and the wall vortices it encounters along the way.

\begin{figure}[htb]
\begin{center}
\scalebox{.4}{\includegraphics{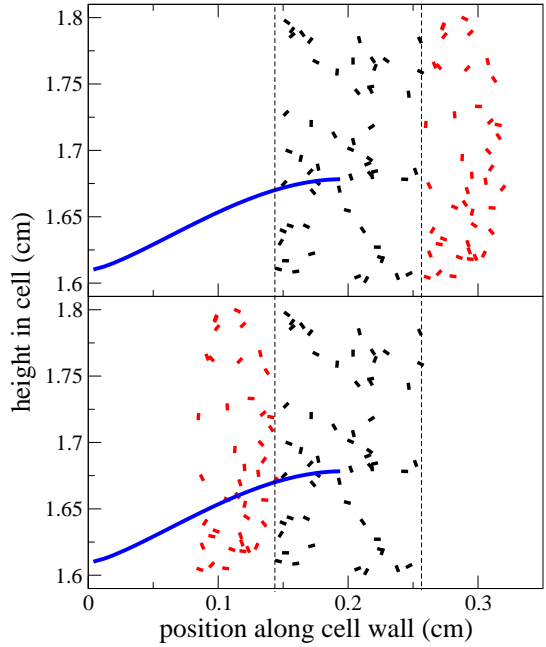}}
\caption{\small  Before moving a mesh patch (top) and after
moving the mesh patch (bottom). The long, moving vortex is on the left in blue.
The patch of mesh that moves, colored red, is to the right of the second vertical
dotted line before moving and left of the first vertical dotted line after
moving. The black vortices between the two vertical dotted
lines do not move in this step.}
\label{f:mvpatch}
\end{center}
\end{figure}

Due to limits in computational power, populating the entire wall of the
cell with mesh vortices is not feasible.  Rather, we define mesh vortices
over a portion of the cell wall near the wall terminus of the moving vortex
line.  The mesh region is an order of magnitude larger than the distances at
which reconnections take place, so pinned vortices outside the region would at
first have no effect under our algorithm.  However, as the macroscopic vortex
precesses, it can leave the initially defined region.  To mimic vortices
covering the entire cell wall, the mesh region must move to track the
precessing vortex.  If both endpoints of a mesh vortex are at least 30\% of
the mesh width behind the moving vortex in the angular direction, we
translate that vortex by exactly the angular width of the mesh to a new
position ahead of the moving vortex.  Conversely, if a mesh vortex is at least
70\% of the mesh width ahead of the moving vortex, we translate that mesh
vortex back to a new position behind the moving vortex.  The asymmetric
conditions adjust for the prevailing motion of the partially trapped vortex.
Similarly, a vortex that lies above (below) the moving vortex by at least half
the mesh height is translated downward (upward) by the mesh height.  Here the
symmetric conditions reflect that the predominant vertical motion comes from
oscillations. Figure \ref{f:mvpatch} illustrates moving the mesh in the angular
direction.  The region of vortices marked in red on the right of
the patch are shifted to the left of the patch at a later time.  In this manner
the moving vortex continues to encounter mesh vortices, but we limit our
calculations to the relevant portions of the cell.  We update the mesh position
every 2000 to 6000 time steps, during which the vortex never traverses more than
30\% of the mesh region.

\begin{figure}[htb]
\begin{center}
\scalebox{.35}{\includegraphics{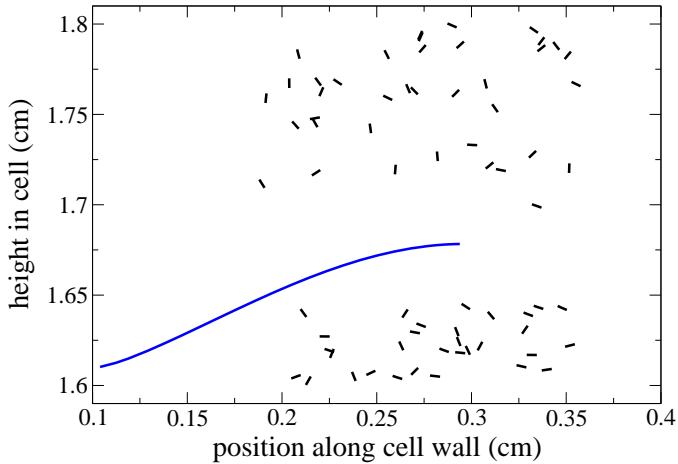}}
\caption{\small  Example of a run where a gap has opened in the
mesh.  The moving vortex is the long solid blue curve extending
from the left into the mesh patch.}
\label{f:badMeshGap}
\end{center}
\end{figure}

Each simulation begins with 40 to 200 mesh vortices.  We create no new mesh
vortices during a simulation, but we remove vortices if their length falls
below a minimum value.  Hence the number of pinned vortices gradually decreases
as the reconnections proceed, and the mesh eventually vanishes. One problem that
sometimes arises is that the mesh develops a gap around the moving vortex, as
shown in Figure \ref{f:badMeshGap}. This typically happens when the mesh patch
covers an insufficient width along the cell wall.  Because of the way we
translate the mesh to keep it centered about the moving vortex, a vortex that
sweeps too fast horizontally relative to its vertical motion can pass through a
given horizontal portion of the mesh multiple times before ever encountering
other portions.  If a gap appears during a run, we discard that run and repeat
with a mesh that extends farther horizontally.

\section*{Results and Discussion}

\begin{figure}[htb]
\begin{center}
\scalebox{.35}{\includegraphics{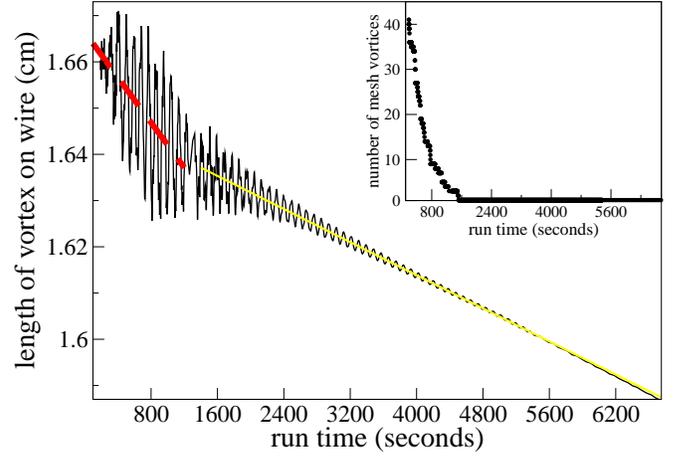}}
\caption{\small  Vertical position of the end of the moving vortex
on the cylinder axis, for a run with $\alpha$ = 0.01. During the first 1000
seconds the vortex interacts with a mesh on the wall. The thick red dashed line 
indicates the average position during this time.  At later times, the
thick yellow solid line gives the average location.  Inset: number of
mesh vortices remaining as a function of time.}
\label{f:twoslopes}
\end{center}
\end{figure}

We find that a vortex mesh can enhance the energy dissipation of a moving vortex
beyond what it would be with no mesh present.  Figure \ref{f:twoslopes} shows
the vertical position of one end of the moving vortex, where it reaches the
cylinder axis. The oscillations are generated by reconnections at the opposite
end of the free vortex.  They continue along the entire vortex line, with
amplitude determined by the friction coefficient $\alpha$.  The dashed red
line for
the first 1200 seconds and the solid yellow line for subsequent times indicate
the average height of the vortex detachment point.  Its decreasing value
indicates that the vertical portion of the vortex shrinks.  Since the portion of
the vortex trapped around the wire is isolated and straight, its kinetic energy
is proportional to its length and the motion of the detachment point indicates
energy loss from the trapped vortex.
The long-time behavior in this run gives a baseline for dissipation
in the absence of mesh vortices. In the initial portion of the simulation, where
the mesh is present, the energy loss is markedly higher.  As seen from the
inset, the mesh disappears shortly before 1600 seconds, coinciding with the
change in the rate of energy loss.

\begin{figure}[htb]
\begin{center}
\scalebox{.35}{\includegraphics{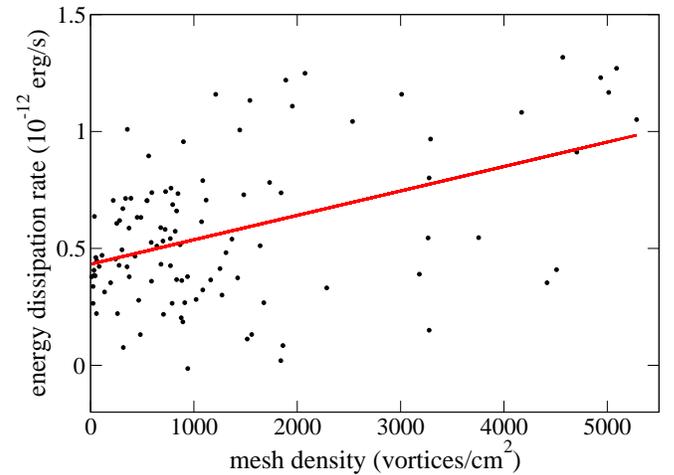}}
\caption{\small Each circle gives average energy loss and average density
for a segment of a run.  See text for further explanation.  The red line 
is a best fit to these points.  Its positive slope shows increase of
the moving vortex's energy dissipation with mesh density.  The data shown here
come from 18 independent runs each using $\alpha$ = 0.01 and beginning with
randomly oriented mesh vortices of length 0.008 cm.}
\label{f:MvsDslope}
\end{center}
\end{figure}

To show more clearly the role of the mesh, we calculate energy dissipation
from the trapped vortex as a
function of the number density of mesh vortices. Time traces, such as that in
Figure \ref{f:twoslopes}, are divided into segments 100 to 300 seconds long, so
that the density is close to constant.  For each segment we calculate the
average values of quantities including number density, energy loss, length of
the moving vortex, and total length of the mesh vortices.  For $\alpha=0.1$ and
$\alpha=0.01$, the time constants for decay for the lowest-frequency Kelvin
oscillations are 120 s and 1200 s, respectively, with the faster
oscillations that contribute most to energy dissipation decaying even more
rapidly.  As a result the behavior during different time segments is relatively
independent.  For $\alpha=0.001$, with a time constant of 12,000 s for the
slowest mode, we instead use only a single time segment near the beginning of
each run, defined by when the number of mesh vortices is between 90\% and 70\%
of the original number.  We select only one segment because the extremely long
time constant means that behavior late in a run depends not only on the
instantaneous mesh density but also on the mesh present earlier in the run. We
combine data from runs with the same friction parameter, mesh vortex length,
and distribution of initial vortex orientations.  Figure \ref{f:MvsDslope} shows
how the energy dissipation rate depends on mesh density, with each point
corresponding to one segment. The dissipation does increase with density,
supporting the idea that reconnections contribute to the energy loss.

Why reconnections increase energy loss from the moving vortex is not obvious.
One possibility is that the reconnections, on average, transfer pieces of the
moving vortex to the wall.  Yet although the length of the moving vortex can
decrease during a reconnection, it can also increase.  During a reconnection
the moving vortex loses the portion between where it intersects the trapped
vortex and the wall.  On the other hand, it adds the portion from one end of
the mesh vortex to the intersection point, as well as adding energy because the
reconnection generally results in high local curvature. Most of our simulations
use straight mesh vortices pulled taut between two pin sites, minimizing
the additional length added to the moving vortex.  Changing the shape of the
mesh vortices does have a significant effect.  We did a few calculations with
semicircular mesh vortices, where the curvature leads to a longer mesh vortex
length between the reconnection site and the wall.  With semicircular vortices,
the moving vortex {\em gains} energy through reconnections. For the physical
experiment, straight vortices seem more plausible.  A pinned vortex can
dissipate energy through oscillations until it arrives at its lowest-energy
configuration, which will usually be close to straight.  If the vortex has too
much curvature, it will encounter the wall during its oscillations and either
annihilate or break into smaller, straighter vortices.

By monitoring the length of the moving vortex immediately before and after a
reconnection, we find that even for straight wall vortices, vortex-vortex
reconnections in our simple reconnection scheme usually lengthen the moving
vortex.  Thus these reconnections do not directly reduce the energy of the
central vortex.  Rather, the energy loss in our
simulations comes from two other sources.  Reconnections between the moving
vortex and the cell wall {\em always} shorten the vortex since a piece remains
behind, attached to the wall.  In addition, the Kelvin waves generated by
reconnections lead to an increase in the
dissipation from mutual friction.  We find that the former mechanism dominates
at low $\alpha$, the latter at high $\alpha$.

\begin{figure}[htb]
\begin{center}
\scalebox{.35}{\includegraphics{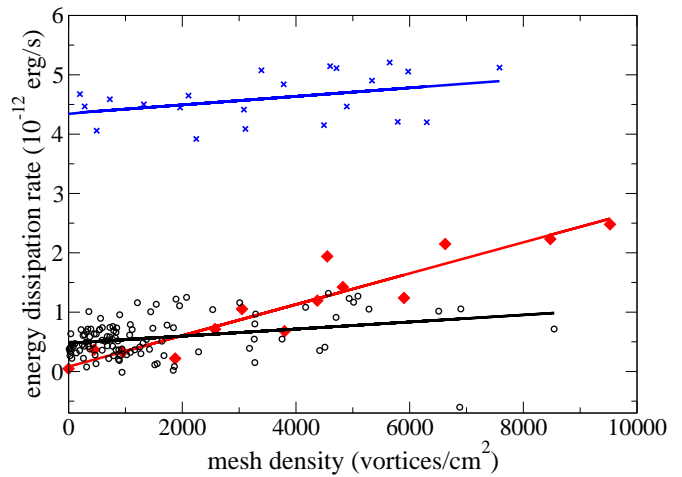}}
\caption{\small  Energy dissipation as a function of mesh
density, for different friction coefficients.  Red filled diamonds:
$\alpha$ = 0.001; black open circles: $\alpha$ = 0.01; blue X's: 
$\alpha$ = 0.1. Lines are best fits to each data set.  The mesh
has a much stronger effect at the lowest friction, and at high mesh density
the dissipation at $\alpha=0.001$ actually exceeds that at $\alpha=0.01$.}
\label{f:MvsDalphaNew}
\end{center}
\end{figure}

The change in mechanism is reflected in Figure \ref{f:MvsDalphaNew}, which
shows energy dissipation rates as a function of mesh density for the three
friction coefficients.  All initial mesh vortices have length 0.008 cm and
random orientations. In each case the dissipation rate increases with mesh
density.  However, the relative influence of the mesh is greater for smaller $\alpha$.
At high enough mesh density even the absolute effect of the mesh is larger
for small $\alpha$; the dissipation for $\alpha = 0.001$ actually {\em
exceeds} that for $\alpha=0.01$. These results for different friction
coefficients are summarized in Table \ref{t:energyloss}; without a mesh the
energy loss comes entirely from mutual friction and scales with $\alpha$, 
but the mesh has more influence as $\alpha$ decreases.

\begin{table}
\begin{tabular}{||c|c|c||}\hline\hline
friction & energy loss & energy loss dependence\\
coefficient  & no mesh present & on mesh density\\
$\alpha$ & ($10^{-12}$ erg/s)& ($10^{-10}$ erg/s per vortex/cm$^2$)\\
\hline\hline
0.1 & 4.35 & 0.717 \\ 
0.01 & .477 & 0.595  \\
0.001 & .0472 & 2.62  \\
\hline\hline
\end{tabular}

\caption{Comparison of energy loss with and without reconnections, 
for several values of friction coefficient. The values
in the second and third columns are the $y$-intercepts and slopes, respectively,
of the best-fit lines in Figure \ref{f:MvsDalphaNew}.}
\label{t:energyloss}
\end{table}

At the higher values of $\alpha$, the pattern of reconnections is typically a
vortex-vortex event followed not long after by a vortex-wall event. The first
reconnection of a run is always vortex-vortex, and we never observe consecutive
vortex-wall reconnections.  Thus the wall reconnections occur only because the
Kelvin oscillations resulting from the original reconnection bring another
piece of the long vortex into proximity with the wall.  The single vortex-wall
event provides no direct explanation for why the mesh increases energy loss; on
average the combination of a vortex-vortex and ensuing vortex-wall reconnection
gives a slight increase in the moving vortex length.  However, the Kelvin waves
produced along the vortex drastically increase its velocity.  Since dissipation
from mutual friction depends on the square of velocity, this has a major effect
on the energy loss.  For straight wall vortices, it more than compensates for
the slight length increase due to reconnections. 

For $\alpha=0.001$, corresponding to a temperature of 850 mK in superfluid
helium, the loss to mutual friction is small even with the induced
oscillations. The low mutual friction means that the Kelvin waves along the
vortex persist for long times, which permits a different form of energy loss. 
Each vortex-vortex reconnection leads to a series of vortex-wall reconnections,
often 100 or more.  As at higher $\alpha$, the vortex mesh initiates
this process; the first reconnection again is {\em always} with another
vortex.  If vortex-wall reconnections outnumber vortex-vortex events even
slightly, the net effect is shortening of the moving vortex; with the large
number of wall events at $\alpha=0.001$, the energy loss from the central
vortex is substantial.  As Figure \ref{f:MvsDalphaNew} shows, this mechanism
which relies on low bulk mutual friction can produce larger energy loss than
exists at higher $\alpha$.

\begin{figure}[htb]
\begin{center}
\scalebox{.35}{\includegraphics{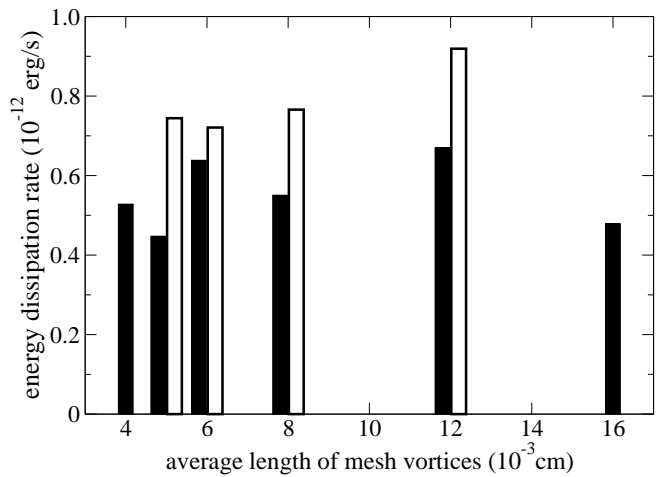}}
\caption{\small Dissipation for simulations with varying mesh vortex lengths.
Greater dissipation is seen in the heterogeneous length mixtures shown as
unshaded columns. Runs with homogenous mesh vortex lengths are shown in black.}
\label{f:MvsLengthBarNew}
\end{center}
\end{figure}

\begin{figure}[thb]
\begin{center}
\scalebox{.35}{\includegraphics{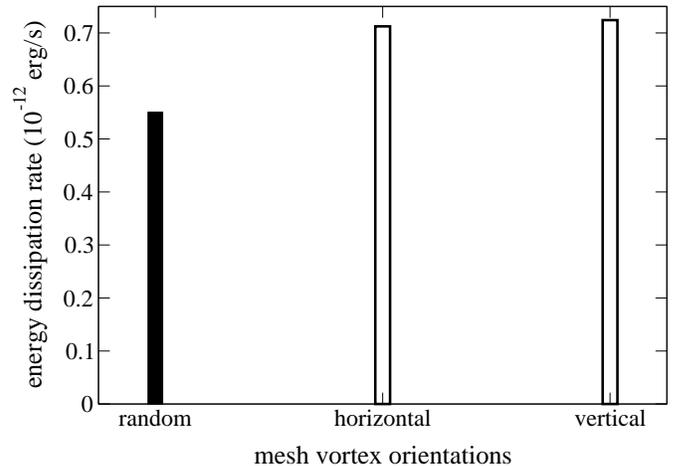}}
\caption{\small Plot of dissipation for simulations with varying mesh vortex
orientations. Greater dissipation is seen the predominantly horizontal  ($\pm
22.5^\circ$) and vertical ($\pm 45^\circ$) textures (shown as empty columns)
than in the one with random orientations (black column).  All data have
$\alpha$ = 0.01 and mesh vortex lengths of 0.008 cm.}
\label{f:MvsOrrsBarNew}
\end{center}
\end{figure}

Other features that the mesh would have in a physical experiment 
appear to enhance the energy loss. Figure \ref{f:MvsLengthBarNew} shows
how energy loss depends on the lengths of individual mesh vortices.  When the
mesh is homogeneous, consisting entirely of vortices of a single length, the
energy loss is independent of the vortex length.  Only the number density of
vortices matters, as is natural if reconnections lead to energy loss mainly
through the Kelvin waves they engender.  As the mesh vortices get longer,
the typical length added to the moving vortex during a reconnection also
increases.  However, this additional length is immediately removed during
the subsequent vortex-wall reconnection.

Heterogeneous mixtures of vortex lengths behave differently.  We use flat
distributions of vortex length in ranges from 0.004 to 0.006 cm, 0.004 to 0.008
cm, 0.004 to 0.012 cm, and 0.008 to 0.016 cm. On comparing runs with a
heterogeneous mesh to those with a homogeneous mesh of the same average vortex
length, we consistently find a higher energy dissipation rate for the
heterogeneous case.  Reconnections occur at a faster rate for the heterogeneous
mesh, with  relatively little length gain but fewer vortex-wall events. The
initial reconnections are with vortices toward the longer end of the range,
which are more likely to intersect the path of the moving vortex. These events
induce Kelvin waves, which make the vortex more likely to encounter the shorter
mesh vortices.  The short vortices have less direct effect on length, but
contribute fully to the Kelvin oscillations along the moving vortex.  For a
friction coefficient of $\alpha=0.01$, where the Kelvin oscillations dominate
energy loss, a heterogeneous mesh dissipates about 50\% more energy than a
homogenous mesh with the same mean vortex length.  A range of vortex lengths is
likely in an experimental setup with irregular wall roughness.

We also consider the orientations of the mesh vortices.   Although we generally
use entirely random orientations, we also test predominantly horizontal
vortices, oriented $\pm 22.5^\circ$ from horizontal in the direction opposite
the free vortex precession, and vertical vortices, directed $\pm 45^\circ$ from
the upward direction.  In each case we take a flat distribution over the
selected range of angles. The average dissipation for the random mesh is 30\%
less than that of the oriented arrangements, as shown in Figure
\ref{f:MvsOrrsBarNew}.  For the horizontal vortices, the length gain from
reconnections with the mesh is shorter than for random vortices.  Figure
\ref{f:reconalg} illustrates this effect.  The red vortex in Figure
\ref{f:reconalg}a represents the moving vortex, which is sweeping
from right to left. The black vortex is nearly horizontal and
directed in the opposite direction, from left to right. Because of the
precession direction and the orientation of the mesh vortices, the vortices meet
near the right end of the mesh vortex, and only a short segment of the mesh
vortex gets incorporated into the moving vortex.  Figure \ref{f:reconalg}d
illustrates the situation after the resulting reconnection. The tendency to add
mainly short pieces accounts for the overall increase in dissipation for this
orientation.  

For predominantly vertical vortices, the increase in dissipation arises because
the vortex-wall reconnections are particularly effective.  The curvature of the
cylinder means that vortices with a large vertical component are closer to the
wall than those that are mainly horizontal.  If a near-vertical mesh vortex is
incorporated into the moving vortex, its proximity to the wall facilitates
subsequent wall reconnections.  The enhanced dissipation from vertical vortices
is particularly relevant because in an experimental setup, vortices created
during rotation are likely to maintain a predominantly vertical orientation when
they partly annihilate at the cell wall after rotation ceases.

Reconnections with pinned vortices have relevance beyond the particular geometry
used here, especially in light of the recent interest in turbulence at very low
temperatures.   In regimes where neither the fluid flow nor temperature can
generate vorticity, remnant vortices have long been seen as a source of the new
vortex lines needed to sustain turbulence \cite{hanninen}. Other experiments
generate turbulence through interactions with solid objects such as oscillating
grids or wires.  Vortices pinned to these structures may play a role in their
large effective mass as they move through superfluid helium
\cite{Charalambous06} as well as directly reducing the onset velocity for turbulence
\cite{Nago}.
Recent work on monitoring flow through tracer particles
\cite{Zhang}
also found that the particles moved much slower than the expected normal fluid
velocity, with the
discrepancy attributed to vortices pinned to the particles \cite{sergeev}
or otherwise distorting their motion \cite{Kivotides}.

\section*{Conclusion}

We find that the interaction of a moving vortex with wall mesh vortices
increases the energy dissipation rate above that observed without this
interaction.  At high temperatures the reconnections generate Kelvin waves,
which increase the vortex velocity and the resulting dissipation from mutual
friction.  At lower temperatures the dominant mechanism is that
Kelvin oscillations
lead to repeated reconnections between the vortex and the wall, which
effectively chop off bits of the moving vortex and transfer them to the wall. 
Our lowest-temperature simulations show the energy loss from the moving vortex
increasing by a factor of 100, and the effect could be even larger at the lower
temperatures achieved in typical experiments.

We thank A. Smith for running some of the early simulations.


\begin{thebibliography}{99}

\bibitem{Donnelly} R.J. Donnelly, {\em Quantized Vortices in Helium II}
(Cambridge University Press, Cambridge, England, 1991).

\bibitem{HV} H.E. Hall and W.F. Vinen, ``Experiments on the propagation of
second sounds in uniformly rotating helium II," {\em Proc. Roy. Soc.} {\bf
A238}, 204-14 (1956); {\em ibid.}, ``The theory of mutual friction
in uniformly rotating helium II," {\em Proc. Roy. Soc.} {\bf A238}, 215-34
(1956).

\bibitem{RR} G.W. Rayfield and F. Reif, ``Quantized vortex rings in superfluid
helium," {\em Phys. Rev.} {\bf 136}, A1194 (1964).

\bibitem{BD} C.F. Barenghi, R.J.Donnelly, and W.F. Vinen, ``Friction on
quantized vortices in helium II," {\em J. Low Temp. Phys.} {\bf 52}, 189-247
(1983).

\bibitem{Tabeling} M. Abid et al., ``Experimental and numerical
investigations of low-temperature superfluid turbulence,'' {\em Euro. J.
Mech. B, Fluids} {\bf 17}, 665 (1998); J. Maurer and P. Tabeling, ``Local
investigation of superfluid turbulence," {\em Europhys. Lett.} {\bf 43}, 29
(1998).

\bibitem{Smith} M.R. Smith, R.J. Donnelly, N. Goldenfeld, and W.F. Vinen,
``Decay of vorticity in homogeneous turbulence,''
{\em Phys. Rev. Lett.} {\bf 71}, 2583 (1993).

\bibitem{Walmsley07} P.M. Walmsley et al.,
``Dissipation of quantum turbulence in the zero temperature limit,"
{\em Phys. Rev. Lett.} {\bf 99}, 265302 (2007); arXiv:0710.1033.

\bibitem{Walmsley08} P.M. Walmsley and A.I. Golov, ``Quantum and quasiclassical
types of superfluid turbulence," {\em Phys. Rev. Lett.} {\bf 100}, 245301
(2008); arXiv:0802.2444.

\bibitem{Tsubota} M. Tsubota, T. Araki, and S.K. Nemirovskii, ``Dynamics of
vortex tangle without mutual friction in superfluid ${}^4$He," {\em Phys. Rev.
B} {\bf 62}, 11751 (2000); arXiv:cond-mat/005280.

\bibitem{Bradley06} D.I. Bradley et al., ``Decay of pure quantum turbulence in 
superfluid 3He-B," {\em Phys. Rev. Lett.} {\bf 96}, 035301
(2006); arXiv:0706.0621.

\bibitem{Vinen01} W.F. Vinen, ``Decay of superfluid turbulence at a very low temperature: The
radiation of sound from a Kelvin wave on a quantized vortex," {\em Phys. Rev. B}
{\bf 64}, 134520 (2001).

\bibitem{Charalambousreview} D. Charalambous et al.,
``Quantum turbulence in ${}^4$He,
oscillating grids, and where do we go next?" {\em J. Low Temp. Phys.} {\bf
145}, 107 (2006).

\bibitem{Donev} L.A.K. Donev, L. Hough, and R.J Zieve, ``Depinning of a
superfluid vortex line by Kelvin waves," {\em Phys. Rev. B} {\bf 64}, 180512(R)
(2001); arXiv:cond-mat/0010240 and L. Hough, L.A.K. Donev, and R.J. Zieve, ``Smooth vortex
precession in superfluid ${}^4$He," {\em Phys. Rev. B} {\bf 65}, 024511 (2001); arXiv:cond-mat/0104525.

\bibitem{Swanson} C.E. Swanson, W.T. Wagner, R.J. Donnelly, and C.F. Barenghi, 
``Calculation of frequency- and velocity-dependent mutual friction parameters
in helium II," {\em J. Low Temp. Phys.} {\bf 66}, 263 (1987).

\bibitem{Adams} P.W. Adams, M. Cieplak, and W.I. Glaberson, ``Spin-up
problem in superfluid ${}^4$He," {\em Phys. Rev. B} {\bf 32}, 171 (1985).

\bibitem{expt} C. Frei, D. Wolfson, and R.J. Zieve, to be published.

\bibitem{Schwarz} K.W. Schwarz, ``Three-dimensional vortex dynamics in
superfluid ${}^4$He: Line-line and line-boundary interactions," {\em Phys. Rev.
B} {\bf 31}, 5782 (1985).

\bibitem{simSamuels} D.C. Samuels, ``Velocity matching and Poiseuille pipe flow
of superfluid helium," {\em Phys. Rev. B} {\bf 46}, 11714 (1992).

\bibitem{simTsubota} M. Tsubota and S. Maekawa, ``Pinning and depinning of two
quantized vortices in superfluid ${}^4$He," {\em Phys. Rev. B} {\bf 47}, 12040
(1993).

\bibitem{simAarts} R.G.K.M. Aarts and A.T.A.M. de Waele, ``Numerical
investigation of the flow properties of He II," {\em Phys. Rev. B} {\bf 50},
10069 (1994).

\bibitem{Aartsthesis} R.G.K.M. Aarts, Ph.D. thesis, Technische Universiteit
Eindhoven (1993).

\bibitem{simqucl} C.F. Barenghi, D.C. Samuels, G.H. Bauer, and R.J. Donnelly,
``Superfluid vortex lines in a model of turbulent flow," {\em Phys. Fluids}
{\bf 9}, 2631 (1997).

\bibitem{lukesim} R.J. Zieve and L.A.K. Donev, ``Stable vortex configurations
in a cylinder,'' {\em J. Low Temp. Phys.} {\bf 121}, 199 (2000); arXiv:cond-mat/0006078.

\bibitem{coeff} From \cite{BD}; our $\alpha$ equals $\gamma/\kappa\rho_s$
from that paper.

\bibitem{recon} J. Koplik and H. Levine, ``Vortex reconnection in superfluid
helium," {\em Phys. Rev. Let.} {\bf 71}, 1375 (1993).

\bibitem{Paoletti} M.S. Paoletti, Michael E. Fisher, K.R. Sreenivasan, and D.P.
Lathrop,  ``Velocity statistics distinguish quantum turbulence from classical
turbulence," {\em Phys. Rev. Lett.} {\bf 101}, 154501 (2008); arXiv:0808.1103 and 
G.P. Bewley, M.S. Paoletti, K.R. Sreenivasan, and D.P. Lathrop,
``Characterization of reconnecting vortices in superfluid helium," {\em
Proc. Nat. Acad. Sci.} {\bf 105}, 13707 (2008); arXiv:0801.2872.

\bibitem{Awschalom} D.D. Awschalom and K.W. Schwarz, ``Observation of a
remanent vortex-line density in superfluid helium," {\em Phys. Rev. Lett.}
{\bf 52}, 49 (1984).

\bibitem{Leadbeater} M. Leadbeater, T. Winiecki, D.C. Samuels, C.F. Barenghi,
and C.S. Adams, ``Sound emission due to superfluid vortex reconnections,"
{\em Phys. Rev. Lett.}{\bf 86}, 1410 (2001); arXiv: cond-mat/0009060.

\bibitem{TT} D.R. Tilley and J. Tilley, {\em Superfluidity and
Superconductivity} (Institute of Physics Publishing, Bristol, 1990), Chapters 1
and 6, p186, 191-3.

\bibitem{Kelvin} W. Thomson, ``Vibrations of a columnar vortex," {\em
Phil. Mag.} {\bf 10}, 155 (1880).

\bibitem{hanninen} R. H\"{a}nninen, M. Tsubota, and W.F. Vinen,
``Generation of turbulence by oscillating structures in superfluid helium
at very low temperatures," {\em Phys. Rev. B} {\bf 75}, 064502 (2007); arXiv:cond-mat/0610224.

\bibitem{Charalambous06} D. Charalambous, L. Skrbek, P.C. Hendry, P.V.E.
McClintock, and W.F. Vinen, ``Experimental investigation of the dynamics of
a vibrating grid in superfluid ${}^4$He over a range of temperatures and
pressures," {\em Phys. Rev. E} {\bf 74}, 036307 (2006).

\bibitem{Nago} Y. Nago et al., 
``Observation of remanent vortices attached to rough
boundaries in superfluid ${}^4$He," {\em J. Low Temp. Phys.} {\bf 158}, 443
(2010).

\bibitem{Zhang} T. Zhang and S.W. Van Sciver, ``The motion of micron-sized particles in He II counterflow as observed by the PIV technique," {\em J. Low Temp. Phys.} {\bf 138},
865 (2005).

\bibitem{sergeev} Y.A. Sergeev, C.F. Barenghi, and D. Kivotides, ``Motion of
micron-size particles in turbulent helium II," {\em
Phys. Rev. B} {\bf 74}, 184506 (2006); 
 {\bf 75}, 019904(E) (2007).

\bibitem{Kivotides} D. Kivotides, ``Motion of a spherical solid particle in thermal counterflow turbulence," {\em Phys. Rev. B} {\bf 77}, 174508 (2008).


\end{thebibliography}
\end{document}